# Sequential Quenching to Predict Semiconductor Defect Concentrations from Formation & Migration Energies: The Case of CdTe:As Doping

*K.A. Arnab[1], I. Chatratin[2], A. Janotti[2], and M.A. Scarpulla[1,3]*

[1.] Department of Materials Science and Engineering, University of Utah, Salt Lake City, Utah 84112, USA

[2.] Department of Materials Science and Engineering, University of Delaware, Newark, Delaware 19716, USA

[3.] Department of Electrical and Computer Engineering, University of Utah, Salt Lake City, Utah 84112, USA

## Abstract

Defect concentrations in semiconductors are strongly influenced by thermal history during growth and cooldown, yet most defect calculations assume either instantaneous quenching from high temperature or that full-equilibrium is maintained - two limiting cases rarely approached in reality. Here, we introduce sequential quenching (SQ) as a 3$^{rd}$ type of defect calculation utilizing defect formation and migration energies to model defect concentrations subject to diffusion-limited kinetics in samples cooled at finite rates. In SQ, the concentration of each defect is frozen at a characteristic temperature determined by its diffusion rate, distance to sources/sinks, and cooling rate. Because different charge-states interact through charge neutrality but freeze at different temperatures, the sequence of freeze-in events is non-commuting. Critically, not all room-temperature SQ solutions can be predicted from full equilibrium (EQ) or full-quenching (FQ) calculations – erroneous predictions are likely without SQ. We illustrate SQ using the example of As-doped CdTe, for which experimental data show differences in doping with cooling rate and between polycrystalline thin-films for photovoltaics and bulk crystals. SQ calculations reveal that fast-diffusing defects such as Cd-interstitials remain mobile to lower temperatures and freeze-in at larger characteristic distances, leading to strong compensation and n-type behavior in rapidly cooled or bulk samples. Slower cooling and reduced characteristic distances suppress donor freeze-in and enhance p-type activation. These results establish SQ as a physically transparent and computationally efficient framework for connecting cooling conditions, sample geometry, and defect kinetics to dopant activation in CdTe and related materials.

*Corresponding Author Email: *mike.scarpulla@utah.edu*

# Introduction

The properties of materials are strongly influenced by the concentrations of point defects and defect complexes, which govern carrier density, compensation, and transport behavior in semiconductors[1–3]. The sensitivity of semiconductors to impurities and native defects at ppm-ppb concentrations is unparalleled in material properties and explains why semiconductor processing is so demanding. Populations of defects are not static; they evolve in response to the thermal and chemical history during crystal growth, irradiation, post-growth thermal processing, and even device operation over long times. Their evolution is governed not only by short-range diffusion and complex formation, but also by long-range transport from/to sources/sinks like dislocations, grain boundaries, surfaces, interfaces, and other extended defects. Therefore, bulk crystals and thin films may exhibit different defect ensembles, as may homoepitaxial thin films vs heteroepitaxial ones having higher dislocation counts. The prime example discussed herein is the widespread observation that efficiently doping CdTe and CdSeTe polycrystalline thin films p-type using group-V elements (grV = N, P, As, Sb) is more challenging than for melt-grown, larger crystals. Another example is that radiation damage in GaN pn diodes grown on freestanding GaN templates can self-anneal at room temperature, while damage was permanent for diodes grown on sapphire with copious threading dislocations. The interpretation is that the dislocations served as sinks for interstitials, preventing Frenkel pair annihilation[4–10].

During crystal growth, local thermal and chemical equilibrium - as for bulk crystallization from the melt - or at least steady-state dynamic equilibrium - as for vapor phase growth processes like molecular beam epitaxy (MBE) or organometallic vapor phase epitaxy (OMVPE) - can be assumed at the growth interface. Usually, the best assumption one can make is that the growth interface and its defect numbers reach full- or dynamic- equilibrium at the highest temperature experienced, since it offers the highest chance of overcoming all kinetic barriers. It is not guaranteed, however, that full chemical and kinetic equilibrium will be reached; for example, it is common to grow structures involving Mg acceptor doping of AlGaInN nitride semiconductors in OMVPE and then to subject them to an activation anneal at a higher temperature in inert gas to remove the passivating H atoms[11–13]. These materials may be thermodynamically unstable at these temperatures in the absence of extreme pressures or reactive nitrogen sources, but the process is optimized to kinetically limit decomposition.

Since diffusion and other kinetic processes become slower as temperature decreases, the next question is "How do details of the cooling process affect defect populations?" Point defects may diffuse to sample surfaces, annihilate each other (as in vacancy-interstitial Frenkel pairs)[10], be generated or annihilated by edge dislocation climb, form complexes like Coulomb-bound donor-acceptor pairs[14,15], condense into clusters (e.g., voids or oxygen-related precipitates in Si[16]), or cross phase boundaries. The extent of all of these processes depends on the time spent at each temperature – the T(t) trajectory. The rates of most kinetic processes are Arrhenian, thus the total extent of processes like diffusion or reaction varies inversely with cooling rate. In principle, each sample having different dimensions and microstructural sources/sinks would require a unique, multidimensional drift-diffusion-Poisson-reaction simulation to predict its final distributions of defects. Such in-depth studies are exactly what is needed to understand specific devices and processes[17–19], but this level of specificity and expensive computation is anathema to making high-level, more generalizable predictions. Additionally, it is rare for all of the needed kinetic parameters and boundary conditions to be available, adding uncertainty to the results. Therefore, a defect calculation approach that approximates defect kinetics and the effects of a coarse-grained description of sample size and microstructure is desired in order to better predict outcomes for a variety of real-world samples.

Herein, we introduce sequential quenching (SQ) as a method for approximating defect concentrations taking into account defect thermodynamics, diffusion kinetics, thermochemical history, and the characteristic distance separating surfaces or other sinks/sources (e.g., crystal size, film thickness, grain size, or distance between dislocations). Conventional calculations of defect concentrations typically assume either full thermodynamic equilibrium (EQ) during infinitely slow cooling or complete freeze-in during instantaneous full quenching (FQ). EQ conceptually corresponds to cooling rate $\gamma=0$ or all kinetic barriers being $<k_BT$ for all temperatures, while FQ corresponds to $\gamma=\infty$ or the limiting kinetic barrier being $>>k_BT$ for all temperatures. Interstitials with low migration barriers maintain EQ even to below room temperature (e.g., Li in Si), while vacancy-mediated diffusers like substitutional dopants frequently exhibit behavior close to FQ. These limiting cases fail to represent real-world growth and processing conditions, for which cooling rates may span several orders of magnitude and defect evolution is governed jointly by thermodynamics, kinetics, and sample geometry[20,21]. Kröger referred to these situations as various types of partial equilibrium – one may assume that a subset of defects has fixed concentrations below threshold temperatures during processing, while others continue to equilibrate based on the thermodynamic conditions. SQ was developed to allow estimation of defect concentrations for finite $\gamma$ between the two typical limits and as a function of characteristic distance.

Figure 1(a) illustrates how the concentrations (denoted by [ ] in the text) of generic neutral defects 1 and 2 would behave during cooling under FQ, SQ, and EQ assumptions. Despite all starting from the same initial conditions at the highest temperature, each scenario results in a unique ensemble of defects at room temperature. SQ is illustrated for both slow and rapid cooling; larger concentrations of each defect (larger supersaturation) are maintained for rapid cooling. Figure 1(b) summarizes the key thermodynamic and kinetic assumptions underlying the three calculation approaches. Under EQ, the cooling behavior of neutral defects is governed only by their own thermodynamics. In EQ, FQ, and SQ cases, we assume that charge carriers equilibrate at all temperatures.

In Fig. 1(a), we illustrated neutral defects which behave independently according to their own thermodynamics and kinetics. Charged defects interact via the charge balance constraint, and thus the results for any charged defect will vary depending on the behaviors of all others and of band carriers. Figure 1(c) illustrates the concentrations vs temperature assuming EQ for a hypothetical semiconductor with one fast-diffusing donor-like (+/0) defect D, one acceptor-like (0/-) defect $A_1$ with high migration barrier, and a $2^{nd}$ acceptor $A_2$ with low migration barriers. Their formation energies are such that, under EQ, intrinsic carriers dominate charge balance below ~2/3 of the max T shown, while the low formation energy of D results in charge balance between D and electrons at high temperatures. However, when considering quenching from intermediate temperatures, we must examine the ordering of defect concentrations.

From 1(c), one could predict that, if the defect system froze in at high temperatures, n-type doping would result, while intrinsic doping would result if the defect system froze in at lower temperatures (since [D], [A1], and [A2] would all be below about 2/3 of the max T scale shown. In contrast, Fig. 1(d) depicts an SQ calculation for the same semiconductor and defects, which predicts p-type doping for finite cooling rate because acceptor $A_1$ with a high migration barrier freezes in at high temperature. Charge balance becomes dominated by the very supersaturated $A_1$ acceptors and holes for all lower temperatures, while $A_2$ and D keep decreasing in number to lower temperatures because of their lower kinetic barriers, and so might only be detectable using capacitive techniques like DLTS. For clarity: in this example, the formation and migration energies follow the ordering $0 < E_{form,D} < E_{form,A1} < E_{form,A2}$ and $0 < E_{mig,D} < E_{mig,A2} < E_{mig,A1}$.

The central result is that sequential freeze-in of different defects (and charge states) produces room-temperature defect ensembles that are not contained within the limiting EQ or FQ solutions, even when all scenarios share the same initial high-temperature state[22]. *In other words, situations will exist in real-world semiconductor processing for which typical defect analysis is completely incapable of predicting the real outcomes.* This is another dramatic illustration that examining defect formation enthalpies and phase equilibrium using density functional theory (DFT) effectively at 0 K is insufficient in terms of predicting real-world behaviors[3,17,18,23–30]. Thus, we assert that SQ is not simply a more complicated analysis that will, in the end, yield marginal additional insight. Rather, combining both defect thermodynamics and kinetics, as in SQ, is critical for accurate predictions.

CdTe is technologically important in bulk crystal form for radiation detectors and as substrates for infrared, but in micron-scale polycrystalline form is also the most commercially-produced thin film photovoltaic technology[8]. Especially in this context, the native defect system tends to compensate for the intentional dopants. To wit, group-V (gr-V = N, P, As, Sb) acceptors can reach activation ratios (p / [grV]) of 50% up to approximately $1-2\times10^{17}$ cm$^{-3}$ in melt-grown bulk crystals, yet it is difficult to achieve few-% activation in films grown by physical vapor deposition techniques suitable for high-throughput solar cell manufacturing[4,8,31–33]. Doping into the desired $10^{15}$-$10^{17}$ cm$^{-3}$ range using closed space sublimation (CSS) or vapor transport deposition (VTD) may require [As] at the 0.1 at% level[5–7,34]. While the phenomenology of low gr-V doping activation is well-established, the microscopic causes remain unproven. Because native defects or the dopant itself in configurations other than substitutional acceptors might compensate holes, as opposed to some detectable other impurity, hypothesis testing has focused on computation.

Around a decade ago, the best available density functional theory computations using hybrid functionals concluded that $P_{Te}$, $As_{Te}$, and $Sb_{Te}$ acceptors were unstable with respect to a distorted AX donor-like configuration, leading to dopant self-compensation[35–37]. Additionally, various complexes were also proposed[8,25,38,39]. More recent computations taking into account CdTe's large spin-orbit coupling in the valence band predict gr-V on Te acceptor ionization levels in better agreement with experiments, especially for Sb[4,29,40]. Notably, these works predict that AX behavior should not occur, and instead propose donor-like Cd interstitials ($Cd_i$), defect complexes involving As ($As_{Te}$-$Cd_i$) as key contributors to compensation[28,29,41–43] whose populations depend sensitively on thermal history and cooling conditions. $As_{Cd}$ "antisite" donors may also appear in Te-rich cases as discussed herein[29]. In this paper, we explore the implications of combined defect thermodynamics and kinetics on As acceptor doping in CdTe.

In order to allow process- and sample-dependent modelling of defect concentrations without full-blown PDE simulations, we introduce sequential quenching (SQ) as part of KROGER[3,44]. KROGER is a computational framework for defect concentration calculations focusing on merging computed formation energetics with real-world thermochemistry and thermochemical constraints specific to processing methods and finite-temperature effects on defect free energies. In SQ, defect species freeze at characteristic temperatures determined by their migration barriers, the cooling rate, and the characteristic distance, while more mobile defects continue to equilibrate to lower temperatures. This approach captures essential non-equilibrium physics without requiring fully spatially resolved reaction-diffusion simulations, enabling efficient estimation of defect concentrations in finite-sized samples. Especially in the past decade, as the computational cost of computing defect migration barriers from DFT has come down and accuracy has increased, many other works have, in their postprocessing analysis, characteristic temperatures at which individual defects' concentrations become frozen-in based on attempt frequency or diffusion length considerations. Our SQ method generalizes and automates these calculations and allows for cross-coupling of defect diffusivities and concentrations – e.g., diffusion of substitutional impurities will depend on the vacancy concentration and impurities will add "drag" to the vacancies' diffusion.

Our SQ framework implements analysis based on the ratio of the characteristic diffusion length $x = \sqrt{Dt}$ to sample dimensions or inter-sink/source distances. It can thus robustly estimate how defect behavior will change with radius in bulk boules and be different between large crystals, thin films, and polycrystals. The condition of charge neutrality, which arises because the self-energy of a charge density of ppm defects dwarfs defect formation energies for volumes larger than ~1 μm$^3$, mutually couples charged defects. If any defect's concentration freezes in, it will affect the resulting defect ensemble. In this work, we apply SQ to As-doped CdTe as a representative and technologically-relevant case study, demonstrating how cooling rate, characteristic distance, and defect migration kinetics jointly govern compensation and gr-V activation.

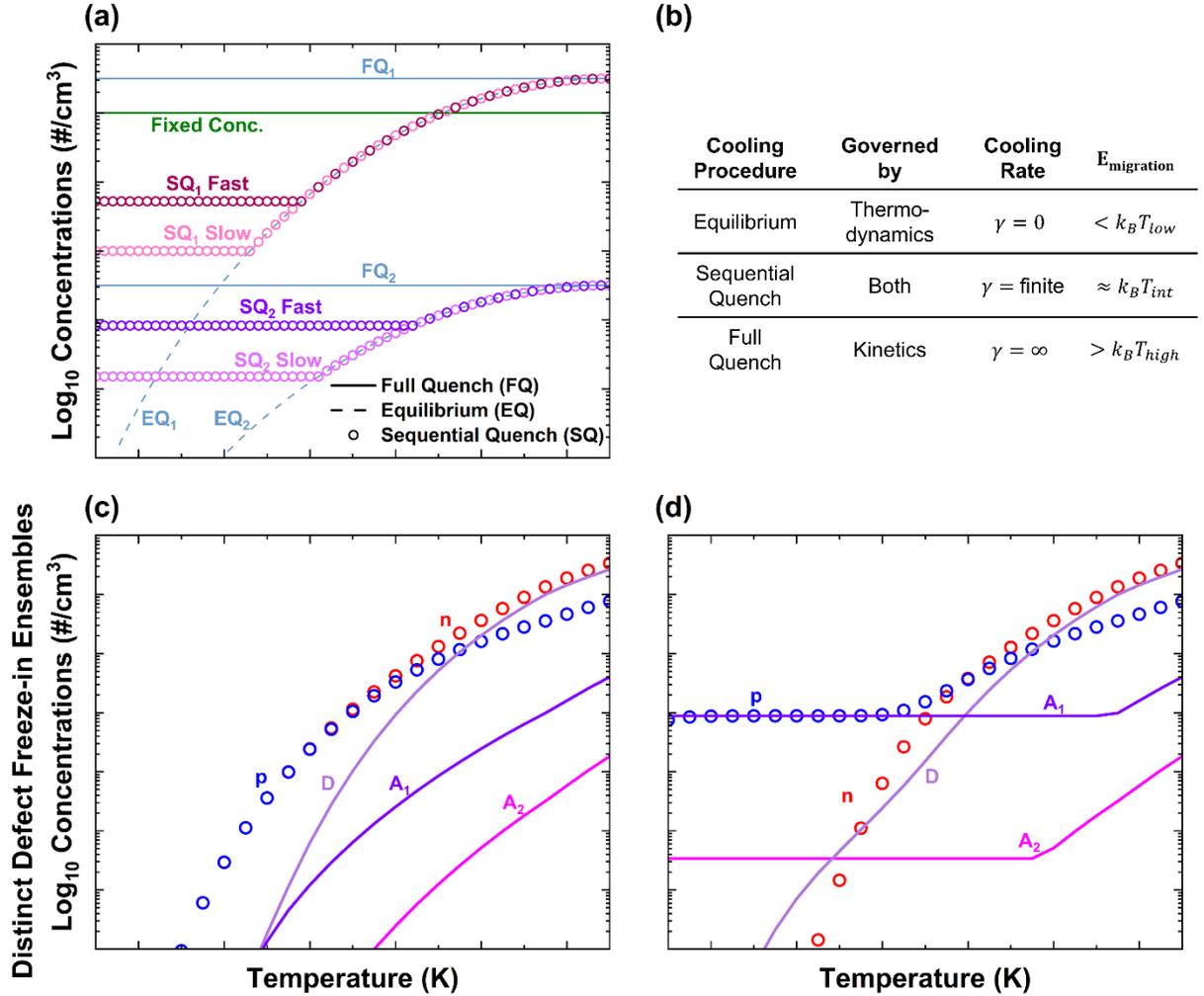

*Figure 1. (a) Concentration-temperature trajectories for generic, uncharged defects 1 and 2 having different formation energies under assumptions of equilibrium (EQ), full quenching (FQ), and sequential quenching (SQ) at slow and fast cooling rates. In SQ, each defect (or charge state) remains equilibrated down to some freezing-in temperature below which the kinetics are too slow to allow further changes. Arrhenius kinetics result in higher freezing temperatures for faster cooling rates (γ). Impurities at concentrations well below solubility limits can be modeled as fixed-concentration defects, which is equivalent to assuming FQ or SQ with $T_{freeze}$ at $T \geq T_{max}$. (b) Table outlining thermodynamic and kinetic aspects of each calculation procedure: SQ uses a unique combination of EQ and FQ for each defect or charge state. (c) Concentrations of donor-like D and acceptor-like $A_1$, $A_2$ defects, as well as band electrons and holes (n, p) for an EQ scenario taking charge balance into account, and in which all defects have equilibrium concentrations less than the intrinsic carrier density $n_i$ at all temperatures. At room temperature, the sample would be intrinsic, meaning that n and p dominate charge balance. (d) As for (c), under an SQ scenario in which $A_1$ freezes in before $A_2$ and D remains able to equilibrate at room temperature (e.g., D might be a mobile interstitial). In (d), the sample is intrinsic at high temperatures, but $A_1$ and p dominate charge balance, and the sample is p-type as a result of $A_1$ freezing in at high temperatures, resulting in large supersaturation.*

Remarkably, the SQ physically-transparent defect analysis framework explains the long-baffling, experimental observations that high p-type doping and activation are straightforward in bulk (>1 cm) crystals, but that polycrystalline thin films with thickness and grain size of a few μm have low activation and hole density. Optimized thermochemical treatments can achieve high activation in epitaxial thin films[34,45]. Especially if $Cd_i$ is taken to be the dominant source of compensation, these observations are difficult to rationalize: wouldn't it be easier for $Cd_i$, which (like many other cations: Cu, Au, Na, Li, etc.) remains mobile at room temperature in the large CdTe lattice, to escape to interfaces and grain boundaries in polycrystalline thin films? The ability of the relatively-computationally-light KROGER with SQ to estimate defect concentrations and predict such differences in defect behaviors as a function of sample geometry and microstructure makes it a unique and powerful tool for defect analysis in specific processing conditions.

## Calculation Methods

Sequential Quenching (SQ) is a calculation procedure enabled by the ability in KROGER[44] to dynamically declare individual charge states, defects, or elements as "frozen" - meaning that their concentrations are fixed for all lower temperatures – or "open" meaning that their numbers continue to change in response to temperature, Fermi level, and relevant element chemical potentials[3]. SQ allows us to account for diffusion-limited kinetics to model defect concentrations resulting from finite cooling rates, rather than the limits of full equilibrium (no kinetic limitations) or full quenching (no diffusion or reaction). The concepts behind the SQ calculation are simple. A set of defects and charge states is defined with formation energies, migration energies, diffusion prefactors, charges, and other relevant attributes. Another attribute for charge states lists the native defect(s) / charge state(s) $X_i$ that mediate its diffusion[14]. At the highest temperature, $T_{max}$, the concentrations of all defects are specified as initial conditions. Herein, we explore both the cases of specified As chemical potential $\mu_{As}$ and specified, fixed total As concentration, but equilibrating all defects through charge balance. However, other initial conditions are possible: for example, concentrations of dopants and Frenkel pairs could be specified in order to predict defect evolution during annealing following ion implantation.

Starting at $T_{max}$, SQ proceeds down a list of temperatures usually ending at $T_{min}$ = 300 K, and at each temperature, the diffusivity $D_j^q$ of each charge state $q$ of defect $j$ is computed from the current concentration(s) of its mediating native defects and its Arrhenius law:

$$D_j^q = D_{j,o}^q \exp\exp\left(-\frac{E_m}{k_B T}\right) = D_{j,oo}^q [X_i] \exp\exp\left(-\frac{E_m}{k_B T}\right) \quad (1)$$

in which $D_{j,oo}^q$ is the diffusion prefactor for the mechanism mediated by defect or charge state $X_i$, and $E_m$ is the corresponding migration energy. If multiple diffusion mechanisms coexist, they are simply added. Throughout, $\gamma \equiv -dT/dt$ denotes the cooling rate, and $L$ is the characteristic transport length (taken as half the smallest relevant dimension, e.g., half grain size or half film thickness). A defect (or charge state) is labeled open if its concentration continues to re-equilibrate at the current temperature, and frozen if its concentration is held fixed for all lower temperatures. We define $T_{freeze}$ for a defect/charge state as the temperature below which its remaining diffusion length during cooldown is insufficient to communicate with sources/sinks over distance $L$. All charged defects remain mutually coupled through charge neutrality at every temperature, so freezing one species constrains the entire ensemble. We note that our current treatment does not capture defect reactions such as complex formation, which will be addressed in later work.

Substitutional defects typically have small $D_o$ and large $E_m$, while interstitials have the opposite trends. Especially for continuous cooling, the exponential dependence on $E_m$ dominates the freeze-in behavior, while $D_o$ shifts freeze-in temperatures only weakly. Herein, we adopt $D_{oo} = 0.25$ cm$^2$/s for all defects, which allows the best agreement with the activation data of Nagaoka et al.[6]. This would correspond to assuming the vibrational frequency $\Gamma \approx 6\times10^{13}$ Hz in the microscopic relation $D_o = \Gamma a_o^2$, which is perhaps slightly high. Ideally, this prefactor would be determined individually for each diffusion mechanism, as would the global adjustable parameter f denoting the fraction of bandgap temperature dependence occurring in the conduction band. Because $D_o$ shifts $T_{\text{freeze}}$ logarithmically, our qualitative conclusions are robust to order-of-magnitude variation in $D_o$. We verified that using $D_o/10$ and $10D_o$ changes absolute freeze-in temperatures but preserves the key trends in activation versus $\gamma$ and $L$. Future work can refine mechanism-specific prefactors from phonon/attempt-frequency analysis.

In our implementation of SQ, we assume that a defect freezes-in at the temperature $T_{\text{freeze}}$, at which its root-mean-square diffusion length *for the remainder of the cooling process* falls below the characteristic distance between sources/sinks. Freeze-in means that it could no longer equilibrate its numbers via diffusion to/from sinks for all lower temperatures. Interestingly, closed-form solutions are possible for the integrated diffusion length of Arrhenius diffusion during continuous cooling $x^2 = \int_0^t D(T(\tau))\, d\tau$ for constant cooling rate (linear) and Newtonian (exponential) $T(t)$ trajectories[46,47]. For linear cooling $T(t) = T_{max} - \gamma t$ with $t \in [0, (T_{max}-T_{min})/\gamma]$ the solution is exact, while the integral for $T(t) = T_{min} + (T_{max} - T_{min})\exp(-\gamma t/(T_{max} - T_{min}))$ does not converge on $t \in [0, \infty]$ but the integral using $T(t) \approx T_{max}\exp(-\gamma t/T_{max})$ provides a very good approximation ($D(T(t))$ and thus the integral $x^2(D(T(t)))$ are well approximated, not $T(t)$ itself). For the linear case, carrying out the integration from t=0 to t=$(T_{current} - T_{min})/\gamma$ gives the integrated diffusion length of a defect during cooling from the current temperature $T_{\text{current}}$ to $T_{\text{min}}$ as:

$$x(T_{current}) = \sqrt{\frac{D_0}{\gamma}\left[T_{current}\exp(\Theta_{current}) - T_{min}\exp\exp(\Theta_{min}) + \frac{E_m}{k_B}\left(Ei(\Theta_{current}) - Ei(\Theta_{min})\right)\right]} \quad (2)$$

in which $\Theta_{current} = \frac{-E_m}{k_B T_{current}}$ and $\Theta_{min} = \frac{-E_m}{k_B T_{min}}$. For the case of exponential free cooling, the approximation mentioned yields:

$$x(T_{current}) = \sqrt{\frac{D_0 \cdot T_{current}}{\gamma} \cdot \Gamma(0, \Theta_{current})} \quad (3)$$

The exponential integral $Ei(z)$ herein is defined as, $Ei(z) = -\int_{-z}^{\infty} \frac{e^{-x}}{x}dx = \int_{-\infty}^{z}\frac{e^x}{x}dx$ and $\Gamma(a,z)$ denotes the upper incomplete Gamma function $\Gamma(a,z) = \int_z^{\infty} x^{a-1} e^{-x}\, dx$. Caution is advised in implementing these results in different programming languages because of differing definitions of special functions. In Mathematica Ei[z] = ExpintegralEi[z], in SciPy Ei[z]=expi(z), however, in Matlab we utilize Ei(z) = -real(expint(-z)), which is valid for z real and positive. Also, in Matlab the definitions of the incomplete Gamma functions differ from Mathematica, but luckily, $\Gamma(a,z)$ is equal to -real(expint(-z)) in Matlab for z both real and positive.

In the case of no coupling between the defect of interest and any others, these equations could simply be solved for $T_{\text{freeze}}$. This situation could be found, for example, for interstitials that are neutral or that never become one of the defects dominating charge balance. However, for a general system of defects coupled via charge balance and diffusion mediation, these represent the current best estimates for its maximum diffusion length, assuming that the concentration of any diffusion-mediating defects never increases at temperatures below $T_{current}$. In each SQ calculation, we sweep over multiple cooling rates and characteristic distances, thus, we step the calculation from $T_{max}$ to $T_{min}$ in small steps (e.g., 1 K for rigor) and at each temperature, check which, if any, defects freeze-in for permutations of these two parameters.

Herein, we use the example of CdTe and explore the case where interfaces and grain boundaries are assumed to be the dominant sources/sinks providing, in the absence of 2$^{nd}$ phase precipitation, the dominant pathway for elimination of defects such as Cd$_i$ during cooling with Cd-rich conditions (or Te$_i$ in Te-rich). The characteristic source/sink distances are taken as ½ of the grain size, film thickness, or bulk crystal's smallest dimension. The incorporation of quasichemical rates for reactions like $V_{Cd} + Cd_i = \emptyset$ or $As_{Te}^- + Cd_i^{2+} = (As_{Te}\text{-}Cd_i)^+$ is beyond the scope of the present work, but we account for As$_{Te}$-Cd$_i$ complexes by treating them as a separate defect with a large migration energy.

Each SQ calculation is begun at a specified high temperature, with the user specifying initial concentrations, whether each element should be treated as fixed-total number or fixed-chemical potential, and whether any defects or charge states should be frozen for the entire calculation. The case of the dopant concentration [As] being specified (as in the green defect labeled "fixed conc." in Fig. 1(a)) is appropriate for liquid phase crystal growth in which the dopant concentration in the well-mixed liquid is below the solubility limit for the maximum temperature where As might diffuse and is locked-in at time of crystallization. The case in which $\mu_{As}$ is set by equilibrium with Cd$_3$As$_2$, amorphous As$_{Te}$, or As$_2$Te$_3$ is appropriate for cases of high high-doping in which As may precipitate out as 2$^{nd}$ phases[4,6,48].

For clarity, we separate the SQ inputs into (i) thermochemical boundary conditions and finite-temperature band-edge assumptions that determine $\mu_i(T)$ and the charge-neutral Fermi level, and (ii) a defect energetics database (formation energies, transition levels, and migration barriers) that controls equilibrium populations and kinetic freeze-in. We describe (i) first, followed by (ii). As in our prior work on defects in β-Ga$_2$O$_3$, we compute chemical potentials from the best-available thermochemical data optimized for the host Cd-Te binary phase system. We define scenarios of CdTe in equilibrium with elemental Cd or Te, corresponding to gedanken experiments with CdTe touching the element or in equilibrium with its equilibrium vapor pressure. We define the 0 of our electron energy scale as the valence band energy at 0 K, and use a free parameter $f$ to parameterize the fraction of bandgap temperature dependence $E_g(T)$ occurring in the conduction band. Unfortunately, we do not have access to the vibrational free energies of each charge state of each defect, nor are the temperature dependence of charge transition levels of the defects available experimentally at material processing temperatures[49]. Thus, we effectively assume that the charge transition levels of all defects remain fixed and that the band edges move versus temperature. In some sense then, the $f$ parameter is used herein to represent how the temperature dependencies of the dominant defect(s)' charge transition levels move relative to the band edges. In future calculations, incorporating vibrational free energies of all defects, $f$ should be set strictly according to the absolute band edge energies. We also utilize the same approximation for the effects of vibrational entropy on all the charge states of each defect in terms of average mode counting. We define the representative mode frequency for the host from its Debye temperature, and then the -TS$_{vib}$ term for 3 quantum harmonic oscillator modes is added for each atom added (e.g., for interstitials) and 3 subtracted for each atom removed (e.g., for vacancies). This results in no change for substitutionals and antisites, which may explain why we find that As$_{Cd}$ donor-like defects may be significant, whereas other analyses have implicated Cd$_i$ only. We do note that Nagaoka reported XFH data suggesting that As$_{Cd}$ may exist for heavily As-doped samples, although it remains somewhat unclear whether that result might have been related to defect clustering or Cd$_3$As$_2$ precipitation[50]. Additionally, in unpublished NMR studies, Pougue, Nagaoka, Scarpulla, and Rockett found solid-state NMR signatures in similar samples that suggested unique atomic configurations. For these reasons, we present our conclusions herein from the best SQ analysis possible at this time, but anticipate that the doping behavior of gr-V's may still reveal further surprises.

First-principles defect energetics were obtained from density functional theory (DFT)[51,52] calculations performed with the projector augmented-wave (PAW) method[53,54] as implemented in VASP[55,56]. To accurately describe the CdTe band gap and defect energetics, we employed the screened hybrid functional of Heyd–Scuseria–Ernzerhof (HSE)[57,58], including spin–orbit coupling (SOC) in the total energies. Following our established CdTe defect workflow, the fraction of Hartree–Fock exchange was set to 33%, and the plane-wave energy cutoff was 300 eV. With this setup, the calculated equilibrium lattice constant of CdTe is $a$ = 6.545 Å and the band gap is $E_g$ = 1.504 eV, consistent with experimental values[59].

Point defects and defect complexes were modeled in 216-atom supercells. All structures were fully relaxed until residual forces were below 0.02 eV/Å. Brillouin-zone integrations were performed at the $\Gamma$ point; tests with off-$\Gamma$ meshes do not alter the qualitative conclusions for the defect trends considered here.

Defect formation energies were evaluated using the standard formalism[60]. For a defect (or defect complex) $D$ in charge state $q$,

$$E_f[D^q; \varepsilon_F, \{\mu_i\}] = E_{tot}(D^q) - E_{tot}(bulk) - \sum_i n_i \mu_i + q(\varepsilon_F + E_v) + E_{corr}^q \qquad (4)$$

where $E_{tot}(D^q)$ and $E_{tot}$(bulk) are total energies of the defective and pristine supercells, $n_i$ counts atoms of species $i$ added ($n_i > 0$) or removed ($n_i < 0$), and $\mu_i$ are the corresponding chemical potentials constrained by CdTe phase stability. The Fermi level $\varepsilon_F$ is referenced to the valence-band maximum (VBM) $E_v$ of bulk CdTe and varies within the band gap. Finite-size corrections $E_{corr}^q$ account for spurious electrostatic interactions and potential alignment in charged supercells[60]; electrostatic potential alignment between charge states was performed by comparing the average electrostatic potential in a region far from the defect. Thermodynamic charge-transition levels $\varepsilon$(q/q′) were obtained from the crossings of $E_f(D^q)$ and $E_f(D^{q'})$.

Migration barriers were determined using the climbing nudged elastic band (cNEB) method[61,62] with at least six intermediate images between the initial and final configurations. To reduce the computational cost while maintaining consistency with hybrid-functional energetics, minimum-energy paths were first relaxed at the GGA level using PBEsol, followed by single-point HSE total energy calculations for initial image and saddle point to obtain barriers consistent with the HSE defect formation energetics. To reduce the computational cost while maintaining consistency with hybrid-functional energetics, minimum-energy paths were first relaxed at the GGA level using PBEsol and then the final migration energy is obtained using HSE.

## Results & Discussion

Figure 2 compares defect concentrations as a function of temperature obtained from sequential quenching (SQ) and equilibrium (EQ) calculations assuming equilibrium with elemental Cd. Figure 2 (a) shows the calculations performed with a fixed [As] concentration of $10^{17}$ cm$^{-3}$ – corresponding exactly to the Cd-solvent THM growth of Nagaoka and Scarpulla[4–6] - while Fig. 2(b) shows the calculations in which $\mu_{As}$ is determined by equilibrium with Cd$_3$As$_2$. Both SQ scenarios are calculated using a characteristic distance of 1 cm and a cooling rate of 0.05 K/s, which correspond to the growth parameters adopted from Nagaoka et al.[6].

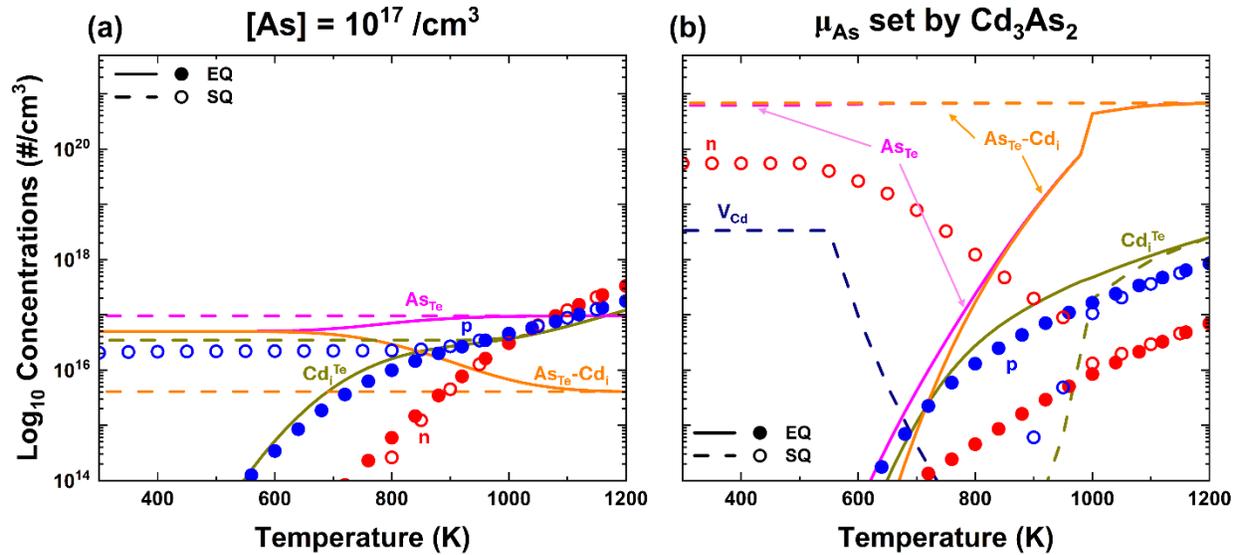

*Figure 2: Defect concentrations as a function of temperature for CdTe equilibrium with Cd, comparing sequential quenching (SQ) for a characteristic distance of 1 cm & γ = 0.05 K/s and equilibrium (EQ) calculations. Two doping scenarios are considered: variable $\mu_{As}$ determined (a) by fixing As concentration of $10^{17}$ cm$^{-3}$, and (b) by equilibrium with the secondary phase $Cd_3As_2$. Solid lines and filled symbols correspond to equilibrium (EQ) calculations, while dashed lines and open symbols correspond to sequential quenching (SQ) calculations.*

In equilibrium calculations (EQ), defect concentrations are allowed to adjust continuously at every temperature, leading to non-monotonic concentration trajectories during cooldown. For example, under equilibrium with Cd conditions, the $As_{Te}$-$Cd_i$ complex remains at concentrations of ~5×10$^{15}$-10$^{16}$ cm$^{-3}$ between 1200 K and ~1000 K, but increases to ~4×10$^{16}$ cm$^{-3}$ at ~700 K till cooling down. This behavior reflects the implicit assumption of equilibrium modeling that defect populations can fully reconfigure even at temperatures where long-range diffusion would be strongly kinetically limited.

In contrast, in sequential quenching (SQ) framework, defects with sufficiently high migration barriers cease to respond directly to further cooling, while more mobile species may continue to evolve. As a result, changes in defect concentrations at lower temperatures can still occur through shifts in charge balance and chemical potentials driven by mobile defects, particularly $Cd_i$ under equilibrium with Cd conditions. This behavior is especially evident under 2$^{nd}$ phase-limited conditions, where variations in $Cd_i$ concentration at low temperature can induce corresponding changes in $V_{Cd}$ populations and defect complexes, even though less mobile defects remain frozen. SQ therefore constrains the active degrees of freedom during cooldown rather than enforcing complete kinetic arrest of all defect populations.

Notably, the qualitative SQ trends are consistent with prior kinetic simulation approaches developed for dopant activation and defect interactions in CdTe (e.g., kinetic models beyond purely thermodynamic treatments)[17,25]. While SQ does not solve spatially resolved reaction–diffusion–Poisson equations, it reproduces the same physically expected directionality: activation decreases when mobile donor-like species remain available to compensate during cooldown, and increases when those species are eliminated or kinetically arrested at higher temperature

The temperature ranges over which individual defects deviate from equilibrium further reflect their relative mobilities. Fast-diffusing species such as $Cd_i$ remain mobile to lower temperatures, whereas defect complexes such as $As_{Te}$-$Cd_i$, having a large migration barrier, freeze in at higher temperatures. Hierarchy of freeze-in behavior is a central feature of SQ and cannot be captured within either equilibrium or full-quench limits.

These differences in defect evolution result in pronounced variations in predicted carrier concentrations. In the fixed-[As] case shown in Fig. 2(a), SQ predicts hole concentrations on the order of ~$10^{13}$ cm$^{-3}$ at room temperature, whereas EQ predicts a much lower value of approximately $10^8$ cm$^{-3}$. This discrepancy arises primarily from differences in the population of $Cd_i$ donors, whose concentration is strongly influenced by the freeze-in of $As_{Te}$-$Cd_i$ complexes at elevated temperatures.

Similar qualitative trends are observed in Fig. 2(b), where the $\mu_{As}$ is constrained by the presence of the $Cd_3As_2$ second phase. EQ predicts that As-related defects diminish rapidly with decreasing temperature and have negligible influence on room-temperature carrier concentrations. In contrast, SQ freezes As-related defects at higher temperatures, where $As_{Te}$-$Cd_i$ dominate compensation. Nevertheless, the presence of a 2$^{nd}$ phase constraint imposes a strong thermodynamic limit on As incorporation, and SQ alone is insufficient to produce p-type behavior under these conditions.

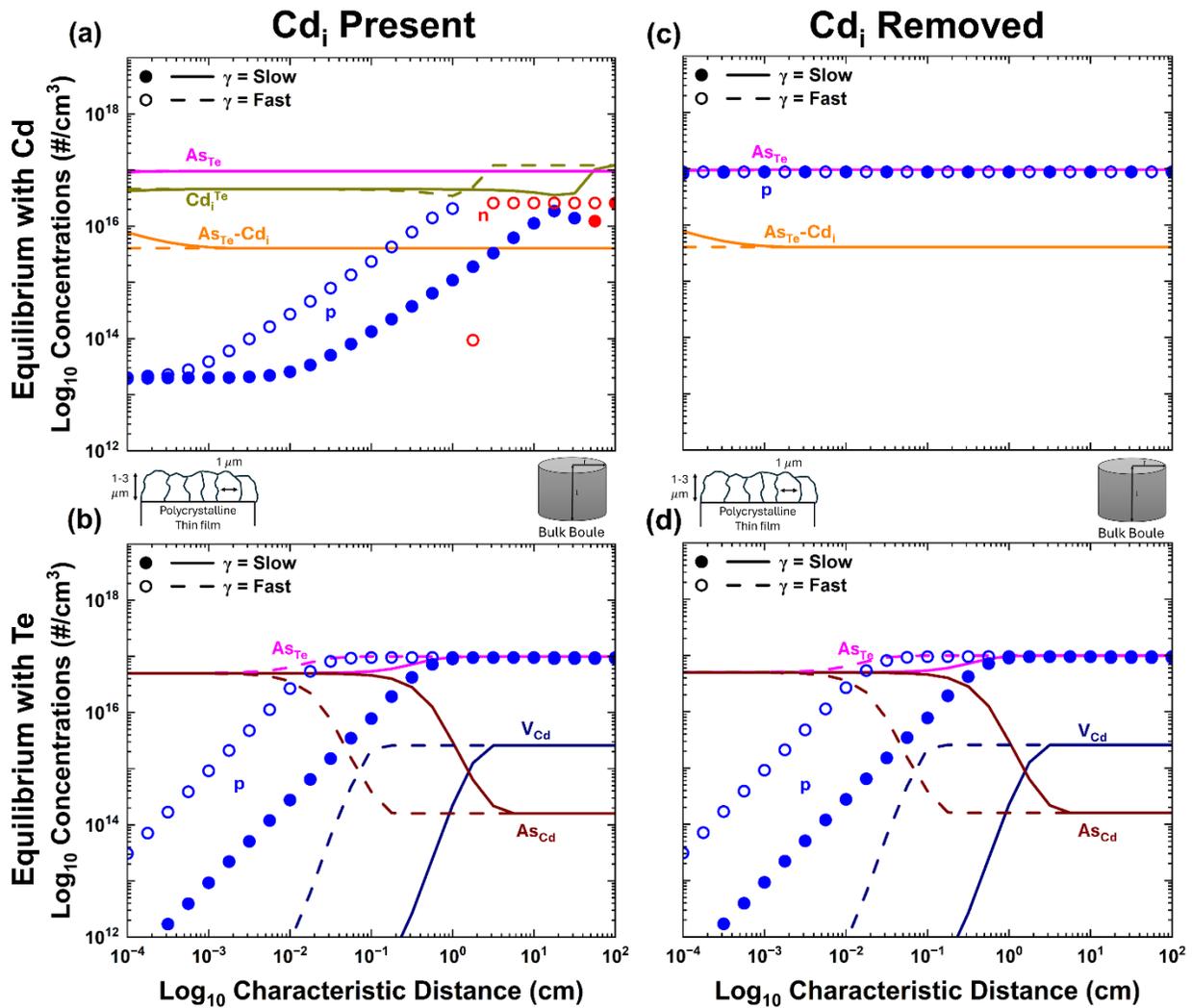

Figure 3: Sequential quenching (SQ) predictions of radial defect and carrier profiles in CdTe under two cooling rates and two thermodynamic boundary conditions. Panels (a) and (b) show SQ results for equilibrium with Cd and Te conditions, respectively,

*for cooling rates of 0.05 K/s (fast cooling; dashed lines and open circles) and $10^{-4}$ K/s (slow cooling; solid lines and filled circles). Faster cooling shifts defect freeze-in to smaller characteristic distances, leading to enhanced donor compensation and the onset of n-type behavior at reduced characteristic distances under equilibrium with Cd conditions.*

*Panels (c) and (d) show the corresponding SQ results after removal of all mobile $Cd_i$ following cooldown, representing the maximum achievable dopant activation when interstitial donors diffuse to sinks during post-growth processing. Under equilibrium with Cd conditions, substantial changes in carrier concentrations are observed due to the dominant compensating role of $Cd_i$, whereas under touching Te conditions, the profiles remain unchanged, reflecting negligible contribution of $Cd_i$ in this regime. In all cases, the total As concentration is fixed at $10^{17}$ $cm^{-3}$.*

Figure 3 (a)-(b) illustrates SQ-predictions of radial defect and carrier profiles for CdTe under two thermodynamic boundary conditions; (a) CdTe in equilibrium with elemental Cd (equilibrium with Cd) and (b) CdTe in equilibrium with elemental Te (equilibrium with Te). For each boundary condition, results are shown for fast (0.05 K/s) and slow ($10^{-4}$ K/s) cooling rates. The total As concentration is held fixed at $10^{17}$ $cm^{-3}$ in all cases to examine the influence of As activation on characteristic distance.

In both boundary conditions, faster cooling shifts defect freeze-in toward smaller characteristic distances, resulting in n-type behavior in equilibrium with Cd scenario beginning at lower characteristic distances (characteristic distance starting n-type behavior changes to ~2 cm from ~50 cm). Increasing characteristic distance further promotes early freeze-in, as longer diffusion lengths are required for defects to reach sinks during cooldown.

Under equilibrium with Cd-rich conditions, $Cd_i$ plays the dominant role in compensation. For fast cooling and large characteristic distances, high concentrations of $Cd_i$ freeze-in (~$10^{17}$ $cm^{-3}$), driving strong donor compensation and n-type behavior. Slower cooling allows $Cd_i$ to remain mobile to lower temperatures, reducing its frozen-in concentration (~$4 \times 10^{16}$ $cm^{-3}$) and thereby decreasing compensation. These results highlight the strong sensitivity of electrical behavior to both cooling rate and sample geometry under Cd-rich boundary conditions.

In contrast, under equilibrium with Te-rich conditions, compensation is governed primarily by $As_{Cd}$, which dominates at small characteristic distances. At larger characteristic distances, the concentration of $As_{Cd}$ remains relatively low concentration of $10^{14}$ $cm^{-3}$, resulting in reduced compensation. At small characteristic distances, however, the $As_{Cd}$ concentration increases to $5 \times 10^{16}$ $cm^{-3}$ and becomes $As_{Cd} = As_{Te}$, leading to strong mutual compensation and a corresponding reduction in hole concentration.

Figures 3(c)-(d) show the corresponding SQ results after modifying the thermodynamic environment following cooldown, allowing mobile $Cd_i$ defects to diffuse out of the sample at low temperatures. This scenario represents an idealized upper bound on dopant activation that may be achieved if free interstitial donors escape to surfaces or grain-boundary sinks during post-growth processing, such as annealing or exposure to oxidizing environments.

Under equilibrium with Cd-rich conditions, removal of $Cd_i$ leads to a substantial reduction in compensation, causing the hole concentration to approach the $As_{Te}$ concentration and yielding activation levels approaching ~98%. Residual $As_{Te}$-$Cd_i$ ($4-8 \times 10^{16}$ $cm^{-3}$) complexes remain frozen in the bulk and maintain a small degree of compensation. In contrast, under equilibrium with Te-rich conditions, the compensation behavior remains unchanged following $Cd_i$ removal, as $As_{Cd}$ is the dominant compensating defect. Consequently, the defect and carrier profiles before and after $Cd_i$ removal are nearly identical in this regime.

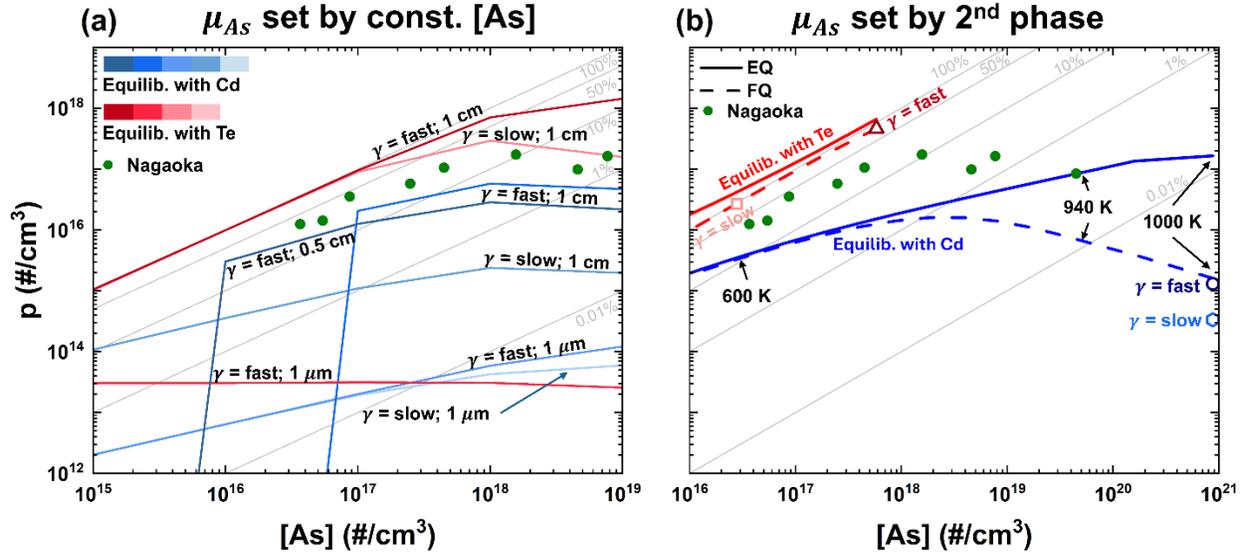

*Figure 4: Predicted As activation in CdTe under different thermodynamic and kinetic modeling scenarios. Panels (a) and (b) show results obtained using a $\mu_{As}$ constrained by a fixed As concentration and by a second phase, respectively. Blue-shaded curves correspond to equilibrium with Cd boundary conditions, while red-shaded curves correspond to equilibrium with Te conditions. Equilibrium with Te conditions consistently yield higher As activation due to the absence of strong interstitial donor compensation, whereas activation under equilibrium with Cd conditions remains significantly limited, with a maximum activation of approximately 30%. For comparison, experimental activation data reported by Nagaoka et al.[6] are included in both panels with olive circles, demonstrating qualitative agreement with the trends predicted by sequential quenching.*

Figure 4 summarizes the predicted As activation in CdTe under different thermodynamic and kinetic modeling scenarios. Figure 4(a) shows results obtained using a fixed As concentration, while Fig. 4(b) shows calculations in which the $\mu_{As}$ is constrained by equilibrium with a second phase ($Cd_3As_2$ and AsTe for the touching Cd and Te conditions, respectively).

In the fixed [As] case shown in Fig. 4(a), sequential quenching (SQ) calculations under equilibrium with Cd-rich conditions yield a maximum activation of approximately 30% for an As concentration of $10^{16}$-$10^{17}$ cm$^{-3}$ at a characteristic distance of 0.5-1 cm under fast cooling. Activation varies systematically with characteristic distance, reflecting the sensitivity of defect freeze-in to diffusion length. For a given characteristic distance, fast cooling generally results in higher activation than slow cooling due to reduced donor equilibration at lower temperatures. In all equilibrium with Cd-rich cases, As activation decreases markedly with increasing dopant concentration. This trend closely resembles experimental activation data reported by Nagaoka *et al.*[6], demonstrating qualitative agreement between SQ predictions and measured behavior in CdTe crystals.

In contrast, equilibrium with Te-rich conditions yield substantially higher activation over a wide concentration range. Near-unity activation is obtained up to approximately $10^{17}$ cm$^{-3}$ [As] doping at large characteristic distances, followed by a decrease in activation at higher dopant concentrations. At large characteristic distances, activation decreases to approximately 2% for slow cooling and 15% for fast cooling at a high As concentration of $10^{19}$ cm$^{-3}$. At small characteristic distances, activation remains low, with maximum values of only a few percent even under fast cooling conditions.

The enhanced activation under equilibrium with Te-rich conditions arises from the absence of Cd$_i$ donors and the relatively low concentration of the dominant compensating defect, As$_{Cd}$. At higher As concentrations, activation decreases due to the stabilization of compensating defects, including Te$_{Cd}$-related defect complexes, which limit the number of electrically active acceptors.

Figure 4(b) shows activation behavior when the $\mu_{As}$ is constrained by equilibrium with a second phase. In this case, equilibrium calculations under equilibrium with Cd conditions exhibit trends similar to experimental data at elevated temperatures. Sequential quenching results under second-phase-limited conditions predict nearly complete activation under equilibrium with Te-rich conditions, whereas activation under equilibrium with Cd-rich conditions remains extremely low for both fast and slow cooling at a characteristic distance of 1 cm. at smaller characteristic distances, activation is strongly suppressed in both boundary conditions, indicating that second-phase constraints impose a dominant thermodynamic limit that cannot be overcome by kinetic freeze-in alone.

The SQ results highlight the central role of Cd$_i$ in controlling compensation under Cd-rich conditions. While Cd$_i$ is highly mobile in the CdTe lattice, its ability to diffuse out of the bulk depends not only on diffusion kinetics but also on the electrostatic landscape within the material.

In polycrystalline CdTe, built-in electric fields associated with grain boundaries and contact depletion regions may significantly influence Cd$_i$ transport. In the dark, depletion regions near grain boundaries and interfaces can create potential barriers that inhibit long-range diffusion of positively charged Cd$_i^{2+}$, leading to accumulation within grain interiors. Under illumination, partial flattening of band bending may temporarily increase the accessible volume for Cd$_i$, allowing redistribution within the CdTe lattice. Notably, the binding energy of the As$_{Te}$– Cd$_i$ complex (0.7 eV)[29] exceeds typical grain-boundary electrostatic barriers (~0.1–0.2 eV)[63], indicating that neither equilibrium band bending nor its partial flattening under illumination is sufficient to dissociate the complex. Consequently, only the population of free Cd$_i$, formed at high temperatures, is expected to participate in redistribution under operating conditions.

Two limiting scenarios may therefore exist for the free Cd$_i$ : (1) Cd$_i$ redistributes reversibly within the CdTe grains with increased volume in response to changing electrostatic conditions, or (2) Cd$_i$ reaches grain boundaries and becomes chemically stabilized or annihilated through reactions with impurities or ambient species. The SQ framework captures the first one through characteristic distances and the second one through the removal of Cd$_i$. This could lead to a pathway for long-term defect evolution that may contribute to light-induced metastability and "wakeup" phenomena observed experimentally.

To our knowledge, no direct experimental evidence of the symmetry-lowering distortion of gr-V acceptors resulting in the AX configuration has been presented to date[64,65]. In both well-known papers claiming to have observed AX centers, the assignment of AX was, in hindsight, based on less-than-definitive evidence. In Ref.[7], the predicted transition energy for the AX center matched an observed 400 meV hole thermal emission energy not present in As-free crystals. The certainty with which this assignment was stated is, in our opinion, not warranted in crystals that were in the same paper reported to contain 2$^{nd}$ phase Cd$_3$As$_2$ precipitates visible with infrared microscopy. Additionally, it is now understood that the computed valence band was misaligned, thus all of the computed transition levels would need to be corrected – meaning that the 400 meV signal would correspond to a different defect. In Ref.[66], the evidence for AX instability was the observation of persistent photoconductivity (PPC). Two quenching-induced doping decay modes were also detected, suggesting multiple defect processes. Photoexcitation of the AX centers to the symmetric substitutional donor would be expected to exhibit PPC, but are not the only possible origins. All that is necessary is that photoexcitation creates a long-lived metastable defect + carrier configuration, which could be a local relaxation as for DX or AX, or a dissociation and reconfiguration of defects or complexes. For example, PPC in BaTiO$_3$ and KTaO$_3$ or photodarkening of Cu-doped β-Ga$_2$O$_3$ are all posited to be caused by hydrogen reconfiguring its location[67–69]. It is important to note that forming complexes like (As$_{Te}$-Cd$_i$)$^+$ from pre-existing As$_{Te}^-$ and Cd$_i^{2+}$ has no impact on net carrier density – the defect quasichemical reaction must also involve carrier capture/emission (similarly for dissociation, whether thermal or photoexcited).

## Conclusions

In this work, we developed a sequential quenching (SQ) framework to model defect evolution under finite cooling rates and applied it to As-doped CdTe. By incorporating defect-specific migration barriers, cooling rate, and characteristic distance to effective sinks, SQ provides a physically grounded description of non-equilibrium defect populations that extends beyond the limiting assumptions of equilibrium and full-quench models. This approach enables the prediction of spatially varying defect and carrier concentrations under experimentally relevant processing conditions.

Application of SQ to CdTe highlights the central role of kinetic limitations in determining dopant activation. Fast-diffusing donor defects, particularly Cd$_i$ remain mobile to low temperatures and can freeze in at large characteristic distances, leading to strong compensation and n-type behavior under equilibrium with Cd-rich conditions. In contrast, slower cooling and reduced characteristic distances limit donor retention and promote higher p-type activation. These results demonstrate that dopant activation in CdTe is governed not only by defect-formation energetics but also by the time-temperature trajectory and the length scales over which mobile defects can diffuse during cooldown.

The analysis further indicates that post-growth redistribution or removal of mobile Cd$_i$ defines an upper bound on achievable activation, while residual defect complexes impose a fundamental compensation limit. In this context, the presence of mobile interstitial donors following growth provides a natural pathway for metastable behavior, where changes in electrostatic conditions, such as those induced by illumination, may enable redistribution of Cd$_i$ without requiring changes in defect formation. Such processes may contribute to experimentally observed "wake-up" phenomena in CdTe, in which carrier concentrations evolve under post-processing conditions.

Taken together, these results establish sequential quenching as a practical and predictive framework for interpreting non-equilibrium defect populations and their evolution under changing thermal and electrostatic environments. More broadly, SQ provides a general methodology for connecting cooling conditions, sample geometry, and defect kinetics to dopant activation and metastability in CdTe and related semiconductor materials

## Conflicts of interest

There are no conflicts to declare.

## Code and Data Availability

The defect properties, including HSE-computed formation energies and migration energies, are available in the Supplementary Materials. The KROGER code is available at
https://github.com/mikescarpulla/KROGER

## Acknowledgement

This work was based [in part] on research sponsored by the U.S. Department of Energy Office of Energy Efficiency and Renewable Energy Solar Energy Technologies Office agreement number 37989 through National Laboratory of the Rockies, operated under Contract No. DE-AC36-08GO28308. IC and AJ were supported by the U.S. Department of Energy, Office of Science, Off*ice of Basic Energy Sciences,* under Award Number DE-SC0025506 and made use of computational resources from the National Energy Research Scientific Computing Center (NERSC), a Department of Energy Office of Science User Facility, through the NERSC award BES-ERCAP 0034471 (m5002), and the DARWIN computing system at the University of Delaware, which is supported by the NSF Grant No.~1919839. The views expressed in the article do not necessarily represent the views of the DOE or the U.S. Government. The U.S. Government retains and the publisher, by accepting the article for publication, acknowledges that the U.S. Government retains a nonexclusive, paid-up, irrevocable, worldwide license to publish or reproduce the published form of this work, or allow others to do so, for U.S. Government purposes.